\documentclass[11pt,preprint]{aastex}
\usepackage{lineno}
\usepackage{epsfig,color}
\usepackage{mathrsfs}
\usepackage{graphicx}
\usepackage{bm}
\usepackage{amsmath}

\def\rblr{$r_{\rm BLR}$}
\def\oiii{[O~{\sc iii}]}
\def\nii{[N~{\sc ii}]}
\def\feii{Fe~{\sc ii}}
\def\hei{He~{\sc i}}
\def\heii{He~{\sc ii}}

\def\Ha{H$\alpha$}
\def\Hb{H$\beta$}
\def\Hc{H$\gamma$}
\def\tHa{$\tau_{\rm{H}\alpha}$}
\def\tHb{$\tau_{\rm{H}\beta}$}
\def\tHc{$\tau_{\rm{H}\gamma}$}
\def\tHei{$\tau_{\rm{He~I}}$}
\def\tH{$\tau_{\rm{H}}$}
\def\tHe{$\tau_{\rm{He}}$}
\def\epshi{$\epsilon_{\rm{H~I}}$}
\def\epshei{$\epsilon_{\rm{He~I}}$}
\def\epsheii{$\epsilon_{\rm{He~II}}$}

\shorttitle{Reverberation Mapping of NGC 2617} \shortauthors{Feng et al.}

\begin{document}

\title{Velocity-resolved Reverberation Mapping of Changing-look AGN NGC 2617}

\author{Hai-Cheng Feng\altaffilmark{1,2,3,\bigstar},
H. T. Liu\altaffilmark{1,3,4, \bigstar},
J. M. Bai\altaffilmark{1,3,4, \bigstar},
Zi-Xu Yang\altaffilmark{2,5},
Chen Hu\altaffilmark{5},
Sha-Sha Li\altaffilmark{2,5},
Sen Yang\altaffilmark{2,5},
Kai-Xing Lu\altaffilmark{1,3,4},
Ming Xiao\altaffilmark{5}}

\altaffiltext{1} {Yunnan Observatories, Chinese Academy of Sciences, Kunming 650011, Yunnan, People's Republic of China}

\altaffiltext{2} {University of Chinese Academy of Sciences, Beijing 100049, People's Republic of China}

\altaffiltext{3} {Key Laboratory for the Structure and Evolution of Celestial Objects, Chinese Academy of Sciences, Kunming 650011, Yunnan, People's Republic of China}

\altaffiltext{4} {Center for Astronomical Mega-Science, Chinese Academy of Sciences, 20A Datun Road, Chaoyang District, Beijing, 100012, People's Republic of China}

\altaffiltext{5} {Key Laboratory for Particle Astrophysics, Institute of High Energy Physics, Chinese Academy of Sciences, 19B Yuquan Road, Beijing, 100049, People's Republic of China}

\altaffiltext{$^{\bigstar}$}{Corresponding authors: Hai-Cheng. Feng, e-mail: hcfeng@ynao.ac.cn, H. T. Liu, e-mail: htliu@ynao.ac.cn, J. M. Bai, e-mail: baijinming@ynao.ac.cn}

\begin{abstract}
  NGC 2617 has attracted a lot of attention after the detection of the changes in spectral type, and its geometry and kinematics of broad-line region (BLR) are still ambiguous. In this paper, we present the high cadence ($\sim$ 2 days) reverberation mapping campaign of NGC 2617 from 2019 October to 2020 May undertaken at Lijiang 2.4 m telescope. For the first time, the velocity-resolved reverberation signature of the object was successfully detected. Both \Ha\ and \Hb\ show an asymmetrical profile with a peak in the velocity-resolved time lags. For each of both lines, the lag of the line core is longer than those of the relevant wings, and the peak of the velocity-resolved lags is slightly blueshifted. These characteristics are not consistent with the theoretical prediction of the inflow, outflow or Keplerian disk model. Our observations give the time lags of \Ha, \Hb, \Hc, and \hei, with a ratio of \tHa:\tHb:\tHc:\tHei = 1.27:1.00:0.89:0.20, which indicates a stratified structure in the BLR of the object. It is the first time that the lags of \Ha\ and \hei\ are obtained. Assuming a virial factor of $f$ = 5.5 for dispersion width of line, the masses of black hole derived from \Ha\ and \Hb\ are $\rm{23.8^{+5.4}_{-2.7}}$ and $\rm{21.1^{+3.8}_{-4.4}} \times 10^{6}M_{\odot}$, respectively. Our observed results indicate the complexity of the BLR of NGC 2617.

\end{abstract}


\section{Introduction}
Active galactic nuclei (AGNs) are powerful extragalactic sources in the universe. In general, AGNs can overwhelm the radiation of their host galaxies within an unresolved spatial region ($<$10$^{-4}$ arcseconds). Depending on the existence of broad emission lines (BELs) in spectra, AGNs are divided into two subclasses: type 1 shows both broad ($>$1000 km s$^{-1}$) and narrow ($<$1000 km s$^{-1}$) emission lines, and type 2 only shows narrow components \citep{KW74}. The BELs are generally believed to be caused by the Doppler motions of gas clouds in broad-line region (BLR). However, the intrinsic differences of the BLR properties between type 1 and type 2 are still under debated. To date, a dozen special AGNs have been detected with the changes in their spectral types on timescales of decades \citep[e.g.,][]{De14,Ki18}, the so-called changing-look AGNs (CL-AGNs), which provide an opportunity to shed light on the mysterious nature of the BLR.

Reverberation mapping (RM) \citep[e.g.,][]{BM82} is a simple and efficient technique for studying the properties of AGNs. It can directly give a reliable mean radius of the BLR (\rblr) via the measurement of the light travel time from ionizing source (tens of gravitational radii) to BLR. The travel time is obtained by calculating the time lag ($\tau$) between the light curves (LCs) of continuum and BEL. Combining the velocity width of BEL ($v$) and \rblr, the black hole virial mass can be estimated as:
\begin{equation}
  M_{\rm{RM}} =f \frac{r_{\rm{BLR}} v^{2}}{G},
\end{equation}
where $f$ is the virial factor, $G$ is the gravitational constant, and \rblr\ = $c \times \tau$ ($c$ is the speed of light) \citep[e.g.,][]{Pe04}. The velocity of BLR cloud is essentially determined by its position and dynamics, and we can therefore separately measure the BEL lags at different velocities to recover the geometry and kinematics of BLR (i.e., the velocity-resolved RM). Although, over 60 AGNs were studied using the spectroscopic RM observations \citep[e.g.,][]{BK15}, the velocity-resolved lags have been successfully reported for only $\sim$20 AGNs due to the limit of spectrum quality \citep[e.g.,][]{De09, Ba11}. There are 4 CL-AGNs in which the lags of \Hb\ have been measured, and only two of them have been probed by the velocity-resolved RM \citep[NGC 3516 and NGC 4151,][]{De18,Fe21}. Thus, it is crucial to extend the number of the velocity-resolved RM campaign for CL-AGNs.

In the context of the traditional unification scheme \citep{An93}, the CL behaviors can be attributed to the motion of dust clouds (the obscuration model). However, the multiwavelength observations of some CL-AGNs challenge the dust occlusion model \cite[e.g.,][]{La15}. Alternatively, significant changes of accretion rate, which may be related to structure changes of accretion disk, are more acceptable for the CL behaviors \cite[e.g.,][]{De14, Sh17, Ro18}. This usually requires a viscus timescale of $\gtrsim 10^{3}$ years. Tidal disruption events (TDEs) are recently adopted to interpret some peculiar objects \cite[e.g.,][]{Me15}. Thanks to the development of time-domain surveys, several transient events accompanied by the type transitions have been detected, and provide fantastic evidence for the TDE scenario. \citet{Ne20} discussed a luminous transient ASASSN-18jd/AT2018bcb in 2MASX J22434289-1659083, sharing similarities with a TDE. \citet{Tr19a} reported the first act of changing phase in 1ES 1927+654 (ASASSN-18el/AT2018zf), and its peak luminosity and timescale are consistent with a TDE model. Nevertheless, the nature of these objects is still unclear. For example, 2MASX J07001137-6602251 (ASASSN-19bt/AT2019ahk) shows a flare at $\sim$ 32 days before appearing to the peak brightness \citep{Ho19a}, while 1ES 1927+654 shows different variability in the optical/ultraviolet (UV) and X-ray bands \citep{Ri20, Ri21}. Interestingly, 1ES 1927+654 only showed a blue continuum at the beginning of the transient, and the BELs appeared a few months later. This indicates that the BLR in 1ES 1927+654 should differ from that of a canonical AGN. \cite{Fe21} also pronounced that there is a unique, elliptical-ring BLR in CL-AGN NGC 3516.

NGC 2617 was classified as a Seyfert 1.8 galaxy in 1994 \citep{Mo96}. In 2013 April, a sudden brightening of the object triggered the alert of All-Sky Automated Survey for Supernovae (ASAS-SN\footnote{http://www.astronomy.ohio-state.edu/asassn}), and the following spectroscopic observations confirmed that NGC 2617 had changed to a Seyfert 1 galaxy \citep{Sh14}. The dramatic variability on such a short timescale cannot be expected from changes in structure of a viscous disk or changes in obscuring material. The properties of its accretion disk have been widely studied \citep[e.g.,][]{Gu17, Fa18}, but only a few studies are related to the BLR. The geometry and kinematics of BLR might be directly related to the physical properties of accretion disk and the radiation pressure on BLR clouds due to irradiation of accretion disk \citep[e.g.,][]{Li17}. Therefore, the studies of BLR might provide some clues to the nature of CL-AGNs. To date, there is only one RM campaign of NGC 2617, carried out in the spring of 2014 (i.e., several months after the CL processes). The campaign reported the mean lags of \Hb, \Hc, and \heii, respectively, suggesting a stratified BLR \citep{Fa17}. In 2016, \citet{Ok17} found a clear new component in the red wing of \Hb. Consequently, NGC 2617 became a double-peaked BEL AGN. The double-peaked profiles of BELs are generally believed to originate from the outer part of an accretion disk \citep[e.g.,][]{St17}. The complexity of the BLR in NGC 2617 is indicated by the ongoing variability of its BELs, and a more detailed study is necessary.

To investigate the properties of the BLRs in CL-AGNs, we started a new RM campaign in 2018. NGC 2617 ($z=0.01421$) is one of the candidates of our RM targets monitored in 2019--2020. The high quality data allow us to measure the velocity-resolved lags of NGC 2617. In this work, we will report the general RM results, and the detailed structure of the BLR will be presented in forthcoming work. We describe the observations and data reduction in Section 2, and present LCs, time series analyses, and measurements of $M_{\rm{RM}}$ in Section 3. In Section 4, we briefly discuss the general RM results, and give the summary. Throughout the paper, we adopt the cosmology with $H_{0}$ = 70 km s$^{-1}$ Mpc$^{-1}$, $\Omega_{\rm{m}}$ = 0.3, and $\Omega_{\Lambda}$ = 0.7.

\section{Observations and Data Reduction}
The observations of NGC 2617 were carried out from 2019 October to 2020 May using 2.4 m telescope at Lijiang Observatory of Yunnan Observatories, Chinese Academy of Science. Following the similar observing procedure of \citet{Fe20}, 75 photometric data points and 63 spectroscopic data points were successfully taken with Yunnan Faint Object Spectrograph and Camera (YFOSC). YFOSC is a versatile CCD with a field of view of 9\farcs6 and a pixel scale of 0\farcs283. The mean seeing condition of 1\farcs15 \citep{Xi20} motivated us to employ a long slit of 2\farcs5. All the spectra were obtained with Grism 3 (a dispersion of 2.93 \AA\ pixel$^{-1}$) and a UV-blocking filter (which cuts off at $\sim$4150 \AA), and we can safely use the data at wavelength of $<$8300 \AA. To improve the accuracy of flux calibration, we simultaneously put the target with a comparison star in the slit. In each night, we also took a spectrophotometric standard star to calibrate the absolute flux of comparison star. The photometric observations were executed with the Johnson $B$ filter in each clear night, and each image was obtained by binning mode (i.e., the CCD bins every 2 pixels in both x and y directions).

The data reduction is the same as in \citet{Fe20}. We first corrected the bias and flat-field for both photometric and spectroscopic images using IRAF software. For the photometry, we extracted the instrumental magnitudes of target and four comparison stars with 15 different apertures. We used the same method as in \citet{Li19} to calibrate the target, and found that the LCs of different apertures are similar to each other. For convenience, the brightness of spectral comparison star was assumed to be 0 mag during the calibration of photometry. We only adopted the results obtained with an aperture radius of 3\farcs4 due to the highest signal-to-noise ratio. The calibrated results are listed in Table 1. For the spectroscopy, the spectra of target and comparison star were extracted from a 42 pixel window ($\sim$11\farcs886) after removing the cosmic-rays. Then, we calibrated the flux of comparison star, and generated a fiducial spectrum by averaging the spectra obtained under good weather conditions. Consequently, an accurate sensitivity function of each exposure can be calculated by the fiducial spectrum. We also corrected the telluric absorption at wavelength of $>$6800 \AA\ using the comparison star. Finally, each spectrum was corrected for the Galactic extinction using the extinction curve of \citet{Fi99}. The values of $E(B - V)$ are obtained from the dust map of \citet{Sc98}. The mean and RMS spectra are shown in Figure 1. Note that the seeing condition will slightly affect the widths of narrow lines \citep{Pe04}, and then it will introduce some residuals in the RMS spectrum. Therefore, we subtracted the fitted narrow components before calculating the RMS spectrum.

\section{Results and Analysis}
\subsection{Host Galaxy and Light Curves}
The host galaxy of NGC 2617 is a clear extended source in our CCD images. Therefore, the effect of seeing condition should be different for AGN and its host galaxy \citep{Fe17,Fe18}, and then should affect the LCs. The brightness of AGN can be well calibrated by the comparison star, while the contribution of the host galaxy is correlated with weather condition. Since the large photometry aperture is used in our data reduction, we did not find any correlation between the brightness and seeing. The slit width (2\farcs5) is much smaller than the photometry aperture size, and there is a large scatter in the LC of 5100 \AA. The broad components are the features of pure AGN, which should be marginally contaminated by the host galaxy. We linearly fitted a continuum under all the BELs, and directly integrated the continuum-subtracted fluxes of \Ha, \Hb, and \Hc. Indeed, the LC scatter of the integrated fluxes of the BELs are much smaller than that of 5100 \AA. However, the some unexpected features of the host galaxy (e.g., absorption lines) can still affect the measurements of the BELs, especially for \hei.

Spectral fitting is a widely used method which can decompose the contributions of host galaxy and AGN \citep[e.g.,][]{Ba15}. The fitting scheme is, with minor exceptions, similar to \citet{Fe21}. Here, we provide a brief summary and describe the differences. The fitting procedure covers the rest-frame wavelength from 4200 \AA\ to 6900 \AA. We masked a narrow spectral window (4250--4450 \AA) around \Hc\ because the broad \Hc\ is overwhelmed by the narrow \Hc\ and \oiii\ $\lambda$4363. The contributions of \feii\ and broad \heii\ are too weak during our observation period, and therefore we ignored them in the fitting. The final fitting model includes: (1) broad \Ha, \Hb, and \hei, (2) several narrow lines and tens of forbidden lines (including coronal lines), (3) a single power-law continuum, and (4) the host template from \citet{BC03}. The broad \Ha\ and \Hb\ are double-Gaussians, while other emission lines are single-Gaussian. We first fitted the mean spectrum with three assumptions: all the narrow lines have the same width, and the ratios of \oiii\ $\lambda\lambda$4959,5007 doublet and \nii\ $\lambda\lambda$6548,6583 doublet are F$_{5007}$/F$_{4959}$ = 3 and F$_{6583}$/F$_{6548}$ = 2.96, respectively. Then, we limited that each individual spectrum has the same spectral index, stellar population of host galaxy, and relative intensity of narrow lines as the mean spectrum. Figure 2 presents the fitted mean spectrum. To evaluate the fitting results, we compared the integrated-flux LCs of \Ha\ and \Hb\ with the fitted results. These two kinds of LCs are consistent with each other, but the scatter of the integrated-flux LCs is slightly larger. We adopted the fitted-flux LCs of \Ha, \Hb, and \hei\ in our analysis. The fitted-flux LC of continuum at 5100 \AA\ is significantly improved, but still with a large scatter. Therefore, only the photometric data are used as the continuum LC. The LC of \Hc\ is obtained from the direct integration. All the LCs are shown in Figure 3, and the corresponding data are listed in Table 1. Their error bars include Poisson errors and systematic errors. The Poisson errors can be obtained from IRAF, while the measurements of systematic errors are different between spectroscopy and photometry. The systematic errors of spectroscopy are calculated via median filter, and the photometric systematic errors are measured from the changes of comparison star \citep[see details in][]{Fe20}. The emission-line errors are dominated by the systematic errors. Thus, we emphasize that the systematic errors obtained from median filter should be overestimated.

\subsection{Time Lags}
The interpolated cross-correlation function \citep[ICCF,][]{WP94} is one of the most robust method for analyzing time series, especially in the case of the well-sampled LCs. Figure 3 shows the results of ICCF between BEL and continuum LCs. The time lag between BEL and continuum variations can be obtained by the location of the peak ($r_{\rm{max}}$) in the ICCF or the centroid of all the points with $r\geq 0.8 r_{\rm{max}}$ in the ICCF. We found that the time lags determined by the peak and centroid, $\tau_{\rm{p}}$ and $\tau_{\rm{c}}$, are consistent with each other. Considering that $\tau_{\rm{p}}$ might be affected by the width of ICCF, the ICCF centroid is adopted as the time lag. Based on the model-independent flux randomization/random subset selection method \citep{Pe04}, a cross-correlation centroid distribution (CCCD) is yielded by $10^4$ realizations of Monte Carlo simulation. The errors of the time lag are given by the 15.87\% and 84.13\% quantiles of the CCCD (equivalent to $\pm 1 \sigma$ if the CCCD is Gaussian). The rest-frame lags and their errors are given in Table 2 for \Ha, \Hb, \Hc, and \hei. The time lags of these BELs show a stratified BLR in NGC 2617.

\subsection{Line Width and Black Hole Mass}
The width of each emission-line can be represented by full width at half maximum (FWHM) or line dispersion ($\sigma_{\rm{line}}$). Traditionally, FWHM and $\sigma_{\rm{line}}$ are directly measured from the mean spectrum or the RMS spectrum, but both of them will be affected by the narrow components. Theoretically, the line width obtained from the RMS spectrum should be less contaminated by the narrow components than that obtained from the mean spectrum. Usually, there are strong narrow-line residuals in the RMS spectrum, depending on spectral quality (accuracy of flux calibration and wavelength calibration) and line spread function. Thus, we only used the fitted broad components to calculate the width of each BEL. \citet{Pe04} found that the fractional errors of black hole mass derived from FWHM are rather large when compared to that derived from $\sigma_{\rm{line}}$. Therefore, we only adopted $\sigma_{\rm{line}}$ as the line-width parameter. We measured $\sigma_{\rm{line}}$ from the mean spectrum after subtracting the fitted narrow components and the host galaxy component. In order to measure the error of each line width, we built up a distribution of $\sigma_{\rm{line}}$ via a bootstrap procedure. Its mean and standard deviation are regarded as the line width and corresponding error, respectively. The intrinsic line widths are estimated by correcting the instrumental resolution ($\sim$1200 km s$^{-1}$ at $\sim$5000 \AA). Combining the measured lags and line widths, the black hole masses can be derived with $f_{\sigma}$ = 5.5 \citep{On04}. The line widths and black hole masses are listed in Table 2.

\subsection{Velocity-resolved Lags}
The high quality data allow us to investigate the velocity-resolved reverberation lags. First, we separately divided \Ha\ and \Hb\ into 11 velocity bins with the same flux in the RMS spectrum. Second, we directly integrated the continuum-subtracted fluxes of each bin, and the resulting LCs are plotted in Figure 4. Then, we measured the time lags and the corresponding errors for \Ha\ and \Hb, as previously described. The velocity-resolved time lags of both \Ha\ and \Hb\ show a single-peak structure (see Figure 5), which are not consistent with the prediction of inflow or outflow. The time lags in the core are longer than those in the wings, but the velocity-resolved time lags do not show symmetrical profiles about their peaks. These trends obviously deviate from the theoretical prediction of the Keplerian disk model. In addition, the peak of velocity-resolved lags is slightly blueshifted. All the observational results imply a complex BLR in NGC 2617. Thus, the detailed geometry is needed to recover the full two-dimensional transfer function \citep[e.g.,][]{Ho04}, and the relevant research will be given in the future work (Feng et al. in preparation).

\section{Discussion and Conclusions}
The sudden brightening of NGC 2617 provides a good opportunity to investigate the properties of CL-AGNs. In general, the CL processes can be interpreted as changes in obscuration or accretion rate. The obscuration model can be well applied to some X-ray selected CL-AGNs \citep[e.g.,][]{Bi05}. The crossing time for an intervening object orbiting outside a BLR can be estimated as \citep{La15}
\begin{equation}
t_{\rm{cross}}= 0.07 \left( \frac{r_{\rm{orb}}}{1 \rm{lt-day}} \right)^{3/2} M_{8}^{-1/2} \arcsin \left( \frac{r_{\rm{src}}}{r_{\rm{orb}}} \right) \/\ \rm{yr},
\end{equation}
where $r_{\rm{orb}}$ is the orbital radius of the foreground object on a circular, Keplerian orbit around the central black hole, $M_{8}$ is the black hole mass in units of $10^8 M_{\odot}$, and $r_{\rm{src}}$ is the true size of the BLR. The above equation is derived by evaluating the time needed for this object to travel the length of an arc that corresponds to the projected size of the BLR. The observational constraints of $r_{\rm{dt}}$ and $r_{\rm{dt}}/$\rblr\ are discussed in estimating $t_{\rm{cross}}$ for the double-peaked BEL CL-AGN NGC 3516, where $r_{\rm{dt}}$ is a dust torus radius \citep[][and references therein]{Fe21}. $r_{\rm{dt}}/$\rblr\ $\sim $ 4--5 is obtained for most reverberation-mapped AGNs with the H$\beta$ BLR and dust torus lag measurements. NGC 3516 has $r_{\rm{dt}}/$\rblr = 7. Thus, it seems appropriate to use $r_{\rm{dt}}/$\rblr = 5 in the CL-AGN NGC 2617. NGC 2617 has averages of $r_{\rm{BLR}}=$ 6.15 lt-days and $M_{8}=0.22$ for H$\alpha$ and H$\beta$, and we obtain $t_{\rm{cross}}$ = 7 yr. NGC 2617 
is Seyfert 2 in 1994 and on 2003 December 30, and Seyfert 1 on 2013 April 25 \citep[see Figure 2 in][]{Sh14}. Our observations show a Seyfert 1 type of optical spectra from 2019 October to 2020 May. The poor sampling of long-term spectroscopic observations makes it impossible to give the reliable durations of the appearance and ``disappearance" of BELs. A long-term monitoring with a good sampling, e.g., twice a year, may give a reliable duration of ``disappearance" of BELs in the CL processes, especially, the continuous and complete process of changing of Seyfert 1 $\rightarrow$ Seyfert 2 $\rightarrow$ Seyfert 1. This will be important to test the obscuration model of CL-AGNs.

On 2013 April 25, NGC 2617 became a Seyfert 1 due to the appearance of optical BELs and a strong continuum blue bump \citep{Sh14}. The appearance of the blue bump implies an increase of accretion rate of the black hole, and consequently the BELs become stronger due to photoionization. The similarity in the X-ray spectral evolution between CL-AGNs and black hole X-ray binaries indicates that the observed CL-AGN phenomena may be related to the state transition in accretion physics \citep{Al20}. The latest research shows that X-ray variability observed in CL-AGNs is produced by changes of accretion disk structure and accretion rate \citep{Ig20}. A natural sequence of the successive weakening of H$\alpha$ and H$\beta$ is produced by the photoionization calculations with the decreasing of ionizing continuum \citep{Gu20}. The luminosity decrease of CL-AGN Seyfert 1 galaxy Mrk 590 is interpreted as an intrinsic change in the mass accretion rate of black hole \citep{De14}. The RM observations of Mrk 590 show $r_{\rm{dt}}(\rm{faint})/r_{\rm{dt}}(\rm{bright})\sim$ 0.3 between the faint and bright phases, where the dust reverberation radius decreased rapidly after the AGN ultraviolet-optical luminosity dropped \citep{KM20}. However, $r_{\rm{orb}}$ in the obscuration model should not vary with the changing accretion rate of the central black hole. The main difference between the obscuration and accretion rate origins of CL-AGNs is likely that $r_{\rm{dt}}$ changes with the AGN ultraviolet-optical luminosity. A photometric and spectroscopic monitoring campaign of $\sim$ 70 days for NGC 2617 from the X-ray through near-infrared (NIR) wavelengths shows that the disk emission lags the X-rays, and the time delays become longer as moving from the UV (2--3 days) to the NIR (6--9 days) \citep{Sh14}. During this monitoring campaign, NGC 2617 went through a dramatic outburst, where its X-ray flux increased by over an order of magnitude followed by an increase of its optical/UV continuum flux by almost an order of magnitude. In this outburst, the X-ray variability unambiguously drives the UV$\rightarrow$NIR variability, almost certainly by irradiating and heating the disk. Thus, the CL behavior in NGC 2617 might be determined by the changes of accretion rate of black hole, switching between different accretion modes.

TDEs can result in the sudden brightening and the slower decaying in the brightness \citep[see][and references therein]{Ho19b}. A TDE is expected to occur every $10^{3}$--$10^{5}$ yr per galaxy \citep{Ma99,Wa04}. The average rates mean only one chance of TDE emerging from the same galaxy in astronomical observations. However,  TDEs strongly prefer certain types of host galaxies, and the TDE rates can be much higher in those galaxies than the average ones, for example, ASASSN-19dj in an extreme post-starburst galaxy KUG 0810+227 \citep[e.g.,][]{Hi21}. Thus, the TDE rates could not seem to give constraints on whether the TDE model works in NGC 2617. \citet{Tr19b} found a new class of flares related to supermassive black hole, in which the spectral features and increased UV flux show little evolution over a period of at least 14 months. This disfavours the tidal disruption of a star as their origin, and instead suggests a longer-term event of intensified accretion. In a TDE regime, the long-term LC should reveal a declining trend of $t^{-5/3}$, and the AGN will quench again in a few years. During our monitoring campaign, about 6.5 years after the outburst in April 2013, NGC 2617 is still in a bright state. The rest-frame 5100 \AA\ continuum luminosity $\lambda L_{5100}= 10^{43.62}$ $\rm{erg \/\ s^{-1}}$, measured from our mean spectrum, is a factor of $\sim$3 brighter than $\lambda L_{5100}= 10^{43.12}$ $\rm{erg \/\ s^{-1}}$ obtained in 2014 \citep{Fa17}. These two brighter states suggest that the optical brightening in NGC 2617 is not from a TDE, and that the changes in its spectral type are not caused by a TDE. The repeatedly brightening processes are likely related to changes of accretion disk structure, i.e., changes of accretion rate. In 2014, a continuum RM campaign gave a lag-wavelength relation of $\tau \propto \lambda^{4/3}$ in NGC 2617 \citep{Fa18}, and this relation implies a standard thin disk. However, the strong hard X-ray emission from NGC 2617 \citep[e.g.,][]{Sh14} requires an additional component than a standard thin disk, such as corona, advection-dominated accretion flow or hot accretion flow \citep[see review by][]{YN14}. \citet{Es97} proposed a two-zone disk model, including an advection-dominated accretion flow within a transition radius $R_{\rm{tr}}$ and an outer standard thin disk, where a change of $R_{\rm{tr}}$ may generate a transition in the accretion regime. The changes of the scale ratio of the standard thin disk to the additional component may lead to the transition of accretion modes.

To date, only one RM campaign about NGC 2617 in 2014 was reported by \citet{Fa17}. Their lags of \Hb, \Hc, and \heii\ show a stratified structure which is consistent with our results, though the emission lines we used are partly different (\Ha, \Hb, \Hc, and \hei). This stratified structure of BLR can be expected from the effects of ionization-energy \citep{CL98} and/or optical depth \citep{KG04}, and is widely detected in AGNs \citep[e.g.,][]{Be10,Zh19,Hu20}. The ionization energies of helium (\epshei = 24.6 eV and \epsheii = 54.4 eV) are much higher than hydrogen (\epshi = 13.6 eV). According to the anti-correlation between gas temperature and radius of BLR, the radius of hydrogen BLR should be larger than that of helium BLR, i.e., \tH\ $>$ \tHe, which is confirmed by our observations (see Table 2). However, in the framework of a simple photoionization model, all the Balmer lines should be expected to originate from the same region, and this expectation is somewhat inconsistent with the observed results of NGC 2617 (see Table 2). This discrepancy can be easily understood after considering the optical depth effect. As the gas densities of BLR clouds increase with decreasing radius, the optical depth of photons at a given frequency should be larger at a smaller radius \citep{Re89}. Hence, the BELs with the larger optical depths at the same radius will appear to show the larger time lags when the BLR clouds at a given radius have the same optical depth for photons at a given frequency. The optical depth of \Ha\ is largest in the same condition, followed by \Hb\ and \Hc. The time lag sequence of \tHa\ $>$ \tHb\ $>$ \tHc\ can be predicted, and this prediction is marginally consistent with our observations (see Table 2). Comparing our results of \Hb\ and \Hc\ to those in \citet{Fa17}, we found that their RM campaign shows the shorter lags (obtained from ICCF) and the broader line widths ($\sigma_{\rm{line}}$). This is a natural outcome of reverberation scenario: when the ionizing continuum flux becomes higher, the BLR clouds at the larger radius will be illuminated, and will yield a larger lag and a narrower profile of BEL.

The symmetric structure of velocity-resolved lags is commonly observed in many AGNs, e.g., NGC 5548 \citep{De09} and Mrk 50 \citep{Ba11}, and is generally regarded as a result of a virialized BLR. The asymmetric structure of velocity-resolved lags is also commonly observed in many AGNs \citep[e.g.,][]{Du18,Zh19}, and is generally regarded as a result of inflow and/or outflow. Our RM campaign gives a slightly asymmetric velocity-resolved signature for NGC 2617 (see Figure 5). The peak of the velocity-resolved lags of \Hb\ locates at the blue side, and the BELs show anomalous profiles. These RM observational features are inconsistent with a simple Keplerian disk-like BLR, and a more sophisticated model is required. Arp 151 shows a similar velocity-resolved RM result \citep{Be08}, and the maximum-entropy method gives a warped disk-like BLR. Based on the same RM campaign of the CL-AGNs NGC 2617 and NGC 3516, the asymmetric structures of velocity-resolved lags were found for the H$\beta$ and H$\alpha$ BELs in NGC 3516, and an elliptical disk-like BLR \citep{Er95} is a possible explanation for the asymmetric structures \citep{Fe21}. The latest numerical simulations show that the eccentricities and orientations of cloud orbits significantly influence the full two-dimensional transfer function of a single disk-like BLR, and the various profiles of BELs can be obtained for different geometries of the single disk-like BLR \citep[see Figures 3 and 7 in][]{KW20}.

In this work, we present a new high cadence RM campaign of CL-AGN NGC 2617 during 2019--2020 with 2.4 m telescope at Lijiang Observatory. The time lags of \tHa = $\rm{6.91^{+1.58}_{-0.78}}$, \tHb = $\rm{5.38^{+0.98}_{-1.12}}$, \tHc = $\rm{4.88^{+2.05}_{-1.94}}$, and \tHei = $\rm{1.04^{+3.24}_{-0.82}}$ days are successfully detected from the high cadence ($\sim$2 days) data, which give \tHa\ and \tHei\ for the first time. The black hole mass of $\sim 2 \times 10^{7}M_{\odot}$ can be derived from both \tHa\ and \tHb. During the period of our observations, NGC 2617 is $\sim 3$ times brighter than $\sim 6.5$ years ago, and then shows the larger lags and the narrower profiles of BELs. A stratified BLR in the object is implied by the lags and velocities of multiple BELs. For the first time, the velocity-resolved reverberation lags of NGC 2617 have been successfully detected. The velocity-resolved lags show a slightly asymmetric structure of \Ha\ and \Hb, and the peak of the velocity-resolved lags of \Hb\ is slightly blueshifted. The lags in the line cores are longer than those in the relevant wings. A complex BLR in NGC 2617 is implied. The accretion rate changes are likely the origin of the CL behavior of NGC 2617.

\acknowledgements {We are grateful to the anonymous referee for constructive comments. We also thanks to Cheng-Liang Jiao, Shi-Yan Zhong, and Yu-Bin Li for the helpful discussion. We also thank the financial support of the National Natural Science Foundation of China (grants No. 11991051, 11703077, and 12073068), the Yunnan Province Foundation (202001AT070069), and the nancial support of the CAS Interdisciplinary Innovation Team. We acknowledge the support of the staff of the Lijiang 2.4 m telescope. Funding for the telescope has been provided by the CAS and the People's Government of Yunnan Province. }

\clearpage

\begin{deluxetable}{ccccccc}
  \tablecolumns{5}
  \setlength{\tabcolsep}{5pt}
  \tablewidth{0pc}
  \tablecaption{LCs of Photometry and Lines}
  \tabletypesize{\scriptsize}
  \tablehead{
  \colhead{JD - 2458000}             &
  \colhead{Mag}                      &
  \colhead{JD - 2458000}             &
  \colhead{\Ha}                &
  \colhead{\Hb}                 &
  \colhead{\Hc}                &
  \colhead{\hei}
} \startdata
770.393164 & 2.402 $\pm$ 0.012 & 779.403218 & 74.67 $\pm$ 1.35 & 11.83 $\pm$ 0.44 & 5.09 $\pm$ 0.57 & 2.91 $\pm$ 0.33 \\
779.397146 & 2.363 $\pm$ 0.007 & 786.387303 & 71.46 $\pm$ 1.35 & 10.68 $\pm$ 0.44 & 4.23 $\pm$ 0.57 & 1.60 $\pm$ 0.32 \\
... & ... & ... & ... & ... \\
\enddata
\tablecomments{\footnotesize The flux of each emission line is in units
of 10$^{-14}$ erg s$^{-1}$ cm$^{-2}$. (This table is available in its entirety
in machine-readable form.)}
\label{Table1}
\end{deluxetable}

\begin{deluxetable}{ccccccc}
  \tablecolumns{6}
  \setlength{\tabcolsep}{5pt}
  \tablewidth{0pc}
  \tablecaption{Rest-frame Lags, Line Widths, and Black Hole Masses}
  \tabletypesize{\scriptsize}
  \tablehead{
  \colhead{Line}                                           &
  \colhead{}                                               &
  \colhead{Lag (days)}                                     &
  \colhead{}                                               &
  \colhead{}{$\sigma_{\rm{line}}$ (km s$^{-1}$)}           &
  \colhead{}                                               &
  \colhead{}{$M_{\rm{RM}}$ ($\times 10^{6}M_{\odot}$)}
} \startdata
\Ha\  & & $\rm{6.91^{+1.58}_{-0.78}}$ & & 1791.7 $\pm$ 4.0  & & $\rm{23.8^{+5.4}_{-2.7}}$    \\

\Hb\  & & $\rm{5.38^{+0.98}_{-1.12}}$ & & 1910.4 $\pm$ 13.0 & & $\rm{21.1^{+3.8}_{-4.4}}$    \\
\Hc\  & & $\rm{4.88^{+2.05}_{-1.94}}$ & & ---               & & ---                           \\

\hei\ & & $\rm{1.04^{+3.24}_{-0.82}}$ & & 2031.4 $\pm$ 45.2 & & $\rm{4.6^{+14.4}_{-3.6}}$     \\
\enddata
\label{Table2}
\end{deluxetable}

\clearpage

\begin{figure*}
  \includegraphics[scale = 0.8]{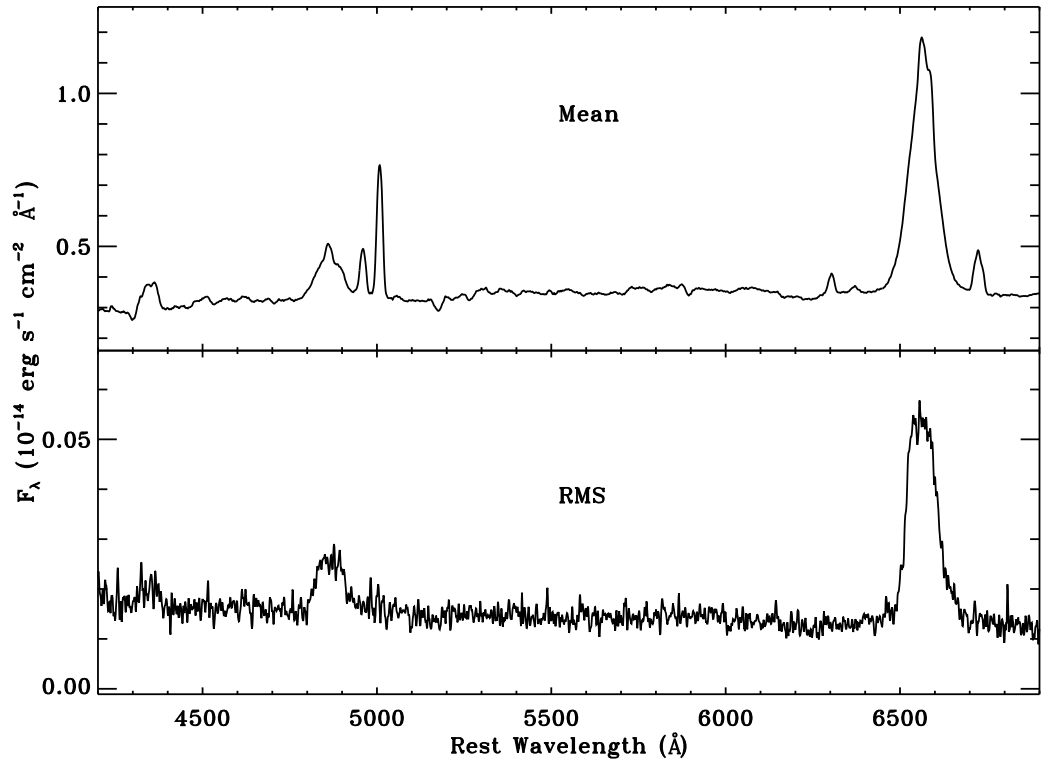}
 \caption{The mean (top) and RMS (bottom) spectra.}
  \label{fig1}
\end{figure*}

\clearpage

\begin{figure*}
  \includegraphics[scale = 0.8]{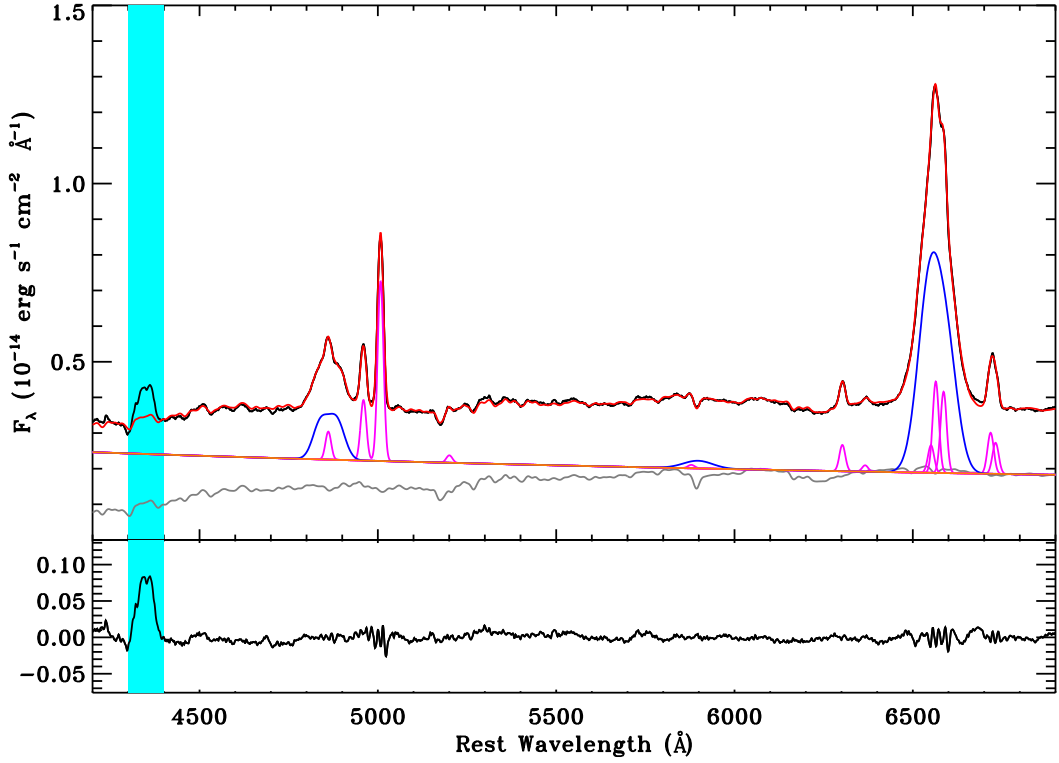}
 \caption{The best-fitting of the mean spectrum and its residual. The top panel shows the observed mean spectrum (black), the best-fitting component (red), host galaxy (grey), power-law continuum (orange), broad emission lines (blue), and narrow emission lines (magenta). The bottom shows the corresponding residuals. The wavelength interval excluded from the fitting is drawn cyan.}
  \label{fig2}
\end{figure*}

\clearpage

\begin{figure*}
 \begin{center}
  \includegraphics[angle=0,scale=0.8]{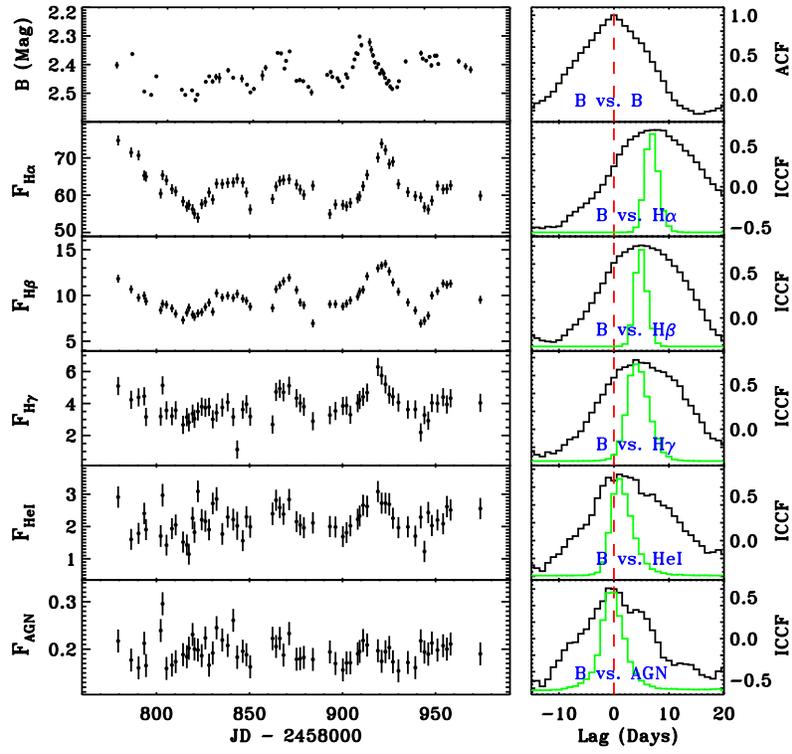}
 \end{center}
 \caption{The left panel shows the LCs of photometry, \Ha, \Hb, \Hc, \hei, and 5100 \AA. The right panel is the ACF (black), CCF (black), and CCCD (green).}
  \label{fig3}
\end{figure*}

\clearpage

\begin{figure*}
 \begin{center}
  \includegraphics[angle=0,scale=1.0]{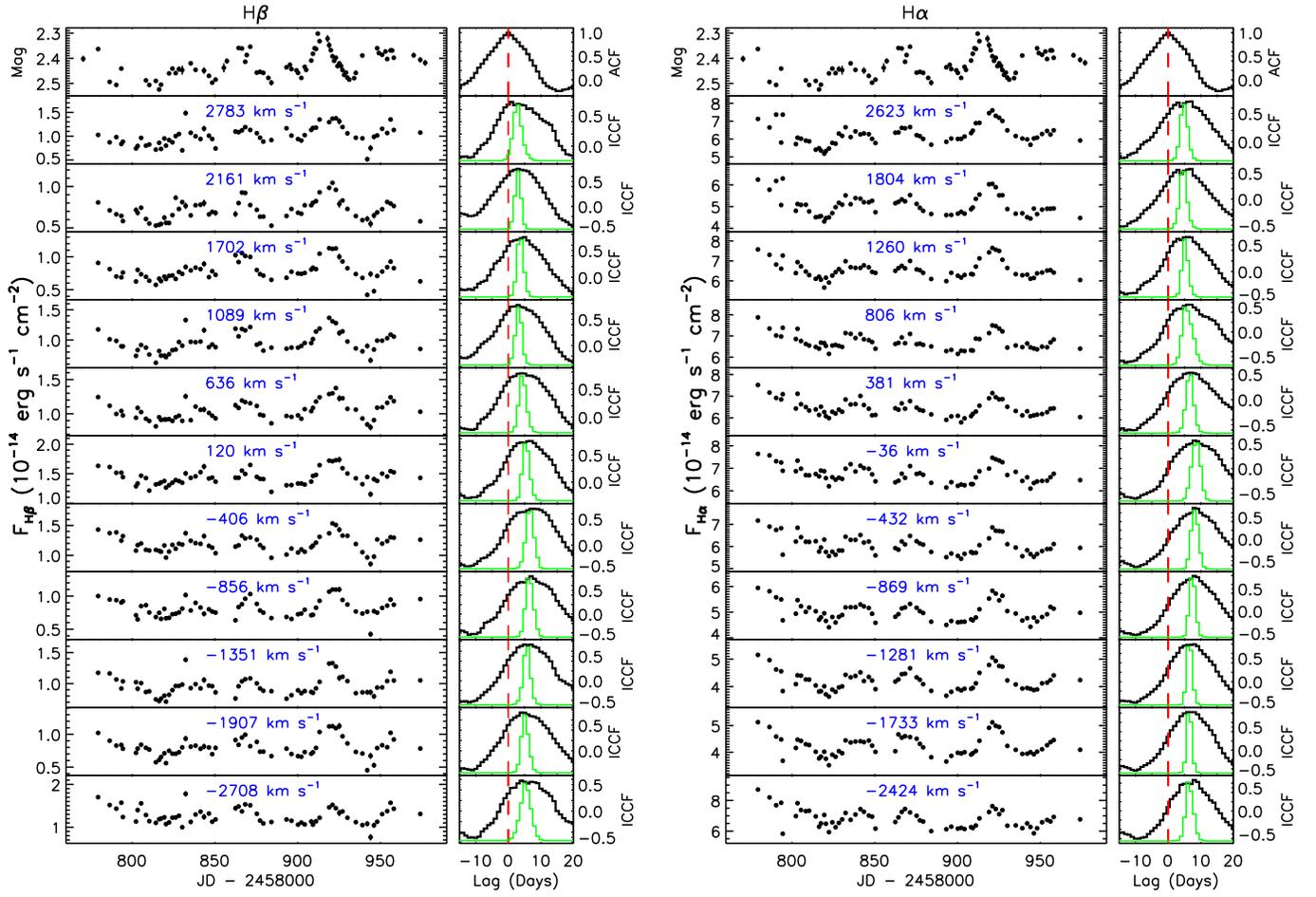}
 \end{center}
 \caption{The LCs in different velocity bins. The format is same as Figure 3.}
  \label{fig4}
\end{figure*}

\begin{figure*}
 \begin{center}
  \includegraphics[angle=0,scale=0.8]{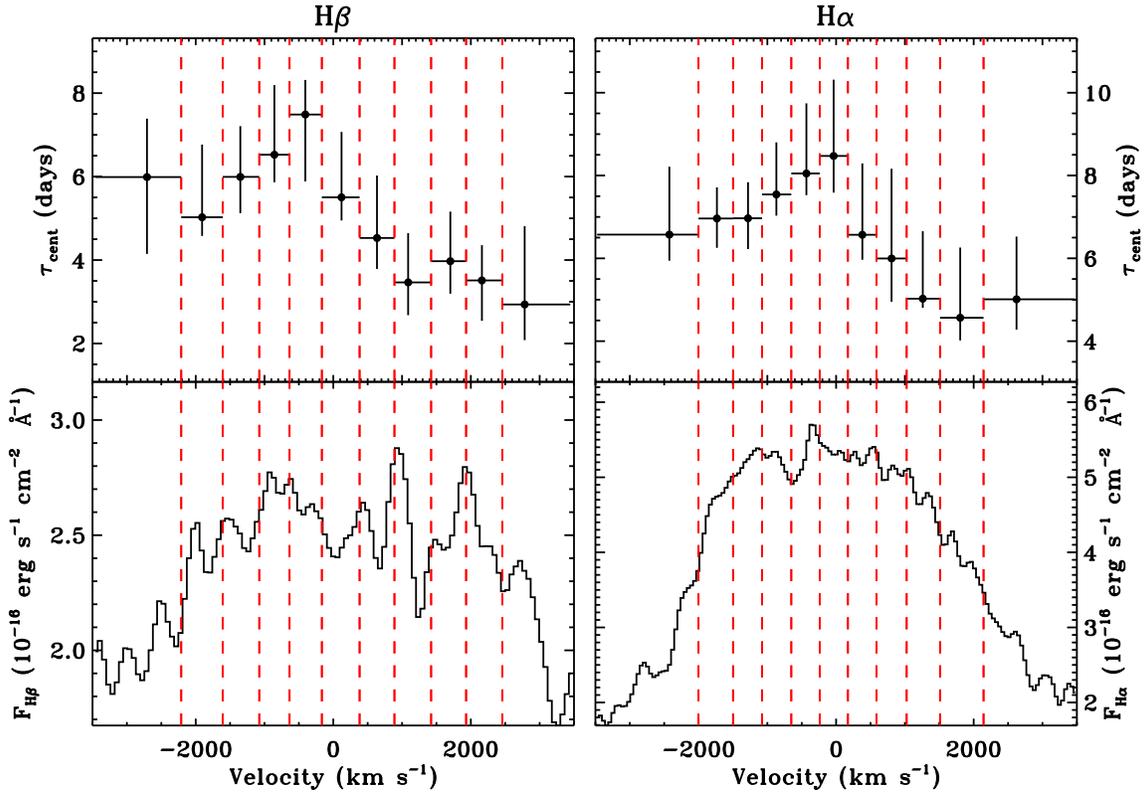}
 \end{center}
 \caption{The top panel shows the velocity-resolved lags of H$\alpha$ (right) and H$\beta$ (left) while the bottom panel shows the RMS spectrum of H$\alpha$ (right) and H$\beta$ (left). The spectral resolution around H$\alpha$ and H$\beta$ are $\sim$ 900 and $\sim$ 1200 km s$^{-1}$, respectively.}
  \label{fig5}
\end{figure*}

\end{document}